\documentstyle[11pt]{article}

\begin{document}
\thispagestyle{empty}
\baselineskip=22pt

\rightline{CLNS-99/1602, KIAS-P99006}
\rightline{hep-th/9901159}
\vskip 2cm

\centerline{\LARGE\bf Membranes from Five-Branes and}
\centerline{\LARGE\bf Fundamental Strings from D$p$ Branes}
\vskip  1cm

\centerline{Piljin Yi \footnote{\tt electronic mail:
piljin@mail.lns.cornell.edu}} 
\vskip 3mm
\centerline{\it Newman Laboratory of Nuclear Studies Cornell University}
\centerline{\it Ithaca, New York 14853-5001, USA} 
\centerline{\it and}
\centerline{\it School of Physics, Korea Institute for Advanced Study} 
\centerline{\it 207-43 Cheongryangri-Dong, Dongdaemun-Gu, Seoul 130-012, Korea}

\vskip 2cm
\centerline{\bf ABSTRACT}
\vskip 5mm
\begin{quote}
We argue that M2 brane is realized as a topological soliton on a coincident 
pair of M5 and anti-M5 branes, as the two five-branes annihilate each other. 
Topology and quantum numbers of this world-volume soliton are discussed in 
some detail, and its formation is explained qualitatively. It follows from a 
compactification that a D4-anti-D4 pair annihilate and produce type II
fundamental strings. The phenomenon is best described as the confinement of
a world-volume $U(1)$ gauge field on D4-anti-D4, where the confined electric 
flux string is identified as the fundamental string. This generalizes to other 
D$p$-anti-D$p$ systems, and solves a puzzle recently pointed out by Witten.
\end{quote}

\newpage
\section{Introduction}

The basic idea behind A. Sen's realization of supersymmetric D branes
from nonsupersymmetric D-brane-anti-D-brane system 
\cite{bound,tachyon,summer}, is simple. The
instability of the initial configuration manifests itself in a
tachyonic mode of open string, which is charged under some world-volume gauge
field. The tachyon rolls down to a true vacuum, but may create a
topological defect in the process. In the simplest case of a D-anti-D pair,
the broken gauge group is $U(1)$, and possible topological solitons
are vortices. A. Sen identified this soliton as supersymmetric D-brane of
two lower dimensions. Recently, this effect was further studied
in a more abstract perspective of K theory  \cite{K,horava,garcia,gukov}.

Physics here is so simple that one naturally wonders whether a similar
phenomenon should exist in other branes, say, NS5 branes and M5 branes. 
Type IIB NS5 branes are related to D5 branes by a U-duality, so whatever 
happens with D5 should happen with IIB NS5 as well. Indeed, world-volume 
theory of type IIB NS5 branes is an ordinary gauge theory, and D3 branes
again appear as vortex solitons. On the other hand,
Type IIA NS5 branes and M5 branes share the same low energy dynamics,
modulo the matter of periodicity of one of the scalar fields. 
We start this paper by trying to understand what happens when a M5 brane
coincides with an anti-M5 brane. 

M5 branes or type IIA NS5 branes admit $(2,0)$ supersymmetry, so their
world-volume degrees of freedom must be in a tensor multiplet, whose
bosonic field content include a chiral 2-form $B^+_{MN}$. Anti-M5
admits anti-chiral 2-form field $B^-$. On the M5-anti-M5 world-volume,
the combined and unrestricted 2-form field $B$ allows a soliton
of co-dimension three that carries a quantized magnetic flux $H=dB$.
This object will be shown to carry membrane charge, and, in the spirit of
Sen's construction, identified as M2.

This construction has an interesting corollary in the context of D branes. 
Note that, compactified on a circle, it reduces
to a system of a D4-anti-D4 pair that annihilates and produces  wrapped 
membranes, or equivalently fundamental strings of type IIA theory. 
Via T-duality transformations, it also tells us that D$p$-anti-D$p$ pairs
annihilate and produce fundamental strings. The physics of the condensation 
is, it turns out, best described as the confinement of the unbroken $U(1)$ 
gauge field on D$p$-anti-D$p$.\footnote{This result may sound quite 
counter-intuitive, because we are used to seeing one of the two $U(1)$'s 
decouple from rest of dynamics. But we need to recall that the decoupling of 
the ``center of mass'' $U(1)$ follows from the combined effect of world-volume 
supersymmetry and translational invariance of string theory. On a 
D$p$-anti-D$p$ pair, supersymmetry is absent. See section 4 for detail.}
The fundamental string is realized as the confined electric flux string of
this $U(1)$ that remains unbroken after the open string tachyon condenses.

This observation solves a puzzle which was recently posed by Witten \cite{K}, 
who observed that the tachyon condensation of Sen breaks only one of two 
$U(1)$ gauge fields on D$p$-anti-D$p$ world-volume, whereas both must 
disappear once the pair annihilate each other. We have just learned that, 
while the second $U(1)$ indeed remains unbroken, it is actually confined
and its only remnant is the fundamental string itself.

The paper is organized as follows. In section 2, we start with a D4-anti-D4
system and explain how it naturally predicts a co-dimension-three  soliton
on M5-anti-M5. The topological coupling that induces the appropriate M2 charge
on such solitons is identified and explained. We also discuss topology of the 
world-volume fields. Section 3 discusses some features of low energy dynamics
that are independent of details of the condensation mechanism, and
explains how the topology of the soliton is supported by the topology
of the would-be Goldstone vector field. Section 4 starts by compactifying
M5-anti-M5 on a circle to deduce that a fundamental string is also
realized on D4-anti-D4, where the condensing tachyonic objects originate
from open D2 branes. These objects are magnetically charged with respect
to the overall $U(1)$ gauge field, as we will explain in detail.
The resulting dual Higgs machanism generalizes to other D$p$-anti-D$p$ systems.
We conclude with a summary.

\section{Membranes from M5-Anti-M5}

Let us start with a system of coincident D4-anti-D4 brane pair, later to be
considered as M5-anti-M5 that are wrapped along the compact circle $S^1$. 
The instability of the configuration 
\cite{T} manifest itself in the lowest lying open string
mode $T$, which becomes tachyonic. From the Chan-Paton factors of 
open strings that connect the two D branes, we know that the tachyon 
is charged with respect to a $U(1)$ gauge field, $A_{12}=A_1-A_2$, 
usually associated with the relative motion of the two branes. 
$A_1$ is the gauge field on D4 while $A_2$ is on anti-D4. Later it will 
turn out that the  overall $U(1)$ gauge
field $A_S\equiv A_1+A_2$ also plays a crucial role, but for the moment
we will concentrate on $A_{12}$. 

The tachyon rolls down to a true vacuum and condenses, $\langle T
\rangle\neq 0$, but in the process its phase may acquire a winding number. 
Otherwise, the final state would correspond to the trivial vacuum of the 
string theory \cite{tachyon}. Because of its coupling to $A_{12}$, the
winding number is accompanied by a localized magnetic flux of $A_{12}=
A_1-A_2$ near $T=0$, and a magnetic vortex soliton is created. On the 
other hand, the world-volume Lagrangian 
contains a Chern-Simons term \cite{DCS} of type,
\begin{equation}
\int_{\rm D4}C_3\wedge dA_1+\int_{\rm anti-D4}C_3\wedge dA_2
=\int_{\rm R^{4+1} }C_3\wedge d(A_1-A_2), \label{CS}
\end{equation}
where $C_3$ is the Ramond-Ramond 3-form tensor field. Note that,
unlike the case of a D4 brane pair where the sum  of gauge fields
couples to $C_3$, the difference of the two gauge field couples to $C_3$. 
This is due to the opposite orientations carried by D4 and anti-D4.

Quantized magnetic flux generated by the winding number of $T$
is localized near $T=0$. Thus,
integrating over the plane transverse to the flux line, the above
coupling reduces to
\begin{equation}
\int_{\{T=0\}\cap R^{4+1} = R^{2+1}}C_3.
\end{equation}
This is precisely how  $C_3$ would have coupled to an ordinary D2 brane 
\cite{D}, from which one concludes that the resulting 
$(2+1)$-dimensional object
must be a D2 brane \cite{summer}. Since D2 brane is the lightest object 
that carries the electric $C_3$ charge, and since the soliton does not seem 
to possess any other spacetime charge, it is a reasonable extrapolation.

Interestingly enough, at least part of spacetime supersymmetry is restored
in the final state, be it the trivial vacuum or D2 branes. 
Although there is no sense in which the soliton was
BPS  with respect to the world-volume theory of D4-anti-D4,
the soliton is actually BPS with respect to the spacetime supersymmetry.
Despite the lack of supersymmetry of the effective field theory that
produces the soliton, one can still reply on the soliton mass formula 
to be stable against continuous change of parameters.

When we regard the D4-anti-D4 pair as a M5-anti-M5 pair wrapped on a compact 
circle $S^1$, the instability should persist. The only difference is that
the tachyonic degrees of freedom must arise 
from open membranes that are stretched
between the five-branes. The boundary of a membrane ending on five-branes is 
string-like, so the object that condenses is a tachyonic world-volume
string. The open string tachyon on the D4-anti-D4 may be regarded as the 
tachyonic string wrapped on $S^1$. Given that D2 branes are nothing 
but transverse M2 branes, and given that the final state is preserved
by restored spacetime supersymmetry, we conclude that a M5-anti-M5 pair 
annihilate and produce M2 branes as world-volume solitons. Let us see this 
phenomenon in M theory setting in more detail.

The M5 brane admits a single chiral 2-form $B^+$ whose 3-form field strength 
$H^+$ is restricted to be self-dual. The anti-M5 brane, on the other hand, is 
identical to the M5 brane, except for the reverse
orientation. The orientation is in turn related to the chirality
of the massless fermions and 2-form tensor. Anti-M5 thus admits anti-chiral
$B^-$, whose field strength  $H^-$ is anti-self-dual. On an M5-anti-M5
system, the two fields combine to become a whole 2-form field, to be denoted
by $B_{MN}$. When we wrap the M5-anti-M5 pair on a circle to make a D4-anti-D4
pair, this $B_{MN}$ reduces to a pair of vector fields. One is obtained
from $B_{5\mu}$, which we may identify with $(A_{12})_\mu$. 
The remaining tensor $B_{\mu\nu}$ is dual to a vector in $(4+1)$ dimensions,
which can be identified with the overall $U(1)$ vector $A_S$.
(It may seem ambiguous which of the two $U(1)$ vectors $A_{12}$ and $A_S$
should be identified with, say, $B_{5\mu}$. But actually, it is simply 
a matter of convention, since, by switching to the dual field strength 
$*H = H^+ - H^-$ in place of $H=H^++H^-$, one effectively exchanges
the longitudinal and the transverse parts.)

Because the 2-form $B$ is now unrestricted, its dynamics can be
described by a Lagrangian. In particular, the Chern-Simon coupling (\ref{CS})
on the D4-anti-D4 pair must come from a similar coupling on the
M5-anti-M5 pair wrapped on $S^1$, and the obvious conclusion we
draw is that there is a term in the world-volume Lagrangian of the form
\cite{CS},
\begin{equation}
\int_{R^{5+1}=\rm M5-anti-M5} C_3\wedge H . \label{ch}
\end{equation}
As we will see in section 4, this coupling is also related to part of
Born-Infeld actions of the D4 brane  and of the anti-D4 brane, upon the 
compactification. Integrating over a localized and quantized 
magnetic flux $H$ on a transverse $R^3$,  we find
\begin{equation}
\int_{R^{2+1}} C_3 .
\end{equation}
Again this is how $C_3$ would have coupled to M2 brane, and so the localized
and quantized magnetic $H$ flux behave as M2 brane. 

So far, we are yet to say how such a soliton of $H$ flux is realized 
dynamically. Since the condensing charged objects are string-like,
one would think that the low energy effective field theory is impossible
to write down. But the essential ingredients of the soliton
construction is not that mysterious. Recall that, when a gauge symmetry 
is broken spontaneously, there always exists an would-be Goldstone mode. 
In the above case of spontaneously broken $U(1)$ gauge theory, it is given by
the phase $\theta$ of $T$, which is eaten up by the gauge field and give 
the latter an effective mass. The effective mass term appears as
\begin{equation}
|T|^2(\partial_\mu\theta-A_\mu)^2,
\end{equation}
and is gauge invariant on its own because a gauge transformation
of $A$ is accompanied by a canceling shift of $\theta$. When the 1-form
gauge field is replaced by 2-form gauge field $B_{MN}$, one also should find a 
similarly gauge-invariant mass term that involves a new would-be Goldstone
mode. A gauge transformation of 2-form $B$ is induced by 1-form, so
the Goldstone boson itself should be 1-form. Call it $G_M$. Thus, the
gauge-invariant mass term must look like \cite{sj},
\begin{equation}
f^2(\partial_M G_N-\partial_N G_M -B_{MN})^2, \label{mass}
\end{equation}
where $f$ is some scalar field that is tachyonic and condenses to restore
spacetime supersymmetry. However complicated is the underlying theory of
M5 branes, this particular feature of condensation should be robust. 
Finally, the would-be
Goldstone boson itself can carry a nontrivial topology, namely the topology
of the complex line bundle associated with the connection $G$.
Significance of this will become obvious in next section.

\section{Condensation of Tachyonic String: a Toy Model}

Let us first recall how one obtains vortex solitons in $U(1)$ gauge theory.
The condensation of charged particle can be treated using a simple
effective field theory of a complex scalar field $T=|T|e^{i\theta}$. 
Let us point out the obvious and say that the phase $\theta $ is periodic 
in $2\pi$. When it couples to a gauge field, the effective Lagrangian is
\begin{equation}
{\cal L} = -\frac{1}{4}{F}_{\mu\nu}^2 +(\partial_\mu |T|)^2
 + |T|^2(\partial_\mu\theta - e{ A}_\mu)^2  - U(|T|) .
\end{equation}
In the decoupled theory, where $e\rightarrow 0$, one sees plane wave of
charged field, $T$, which is promoted to charged particles
upon quantization. In addition there exist some classical solitons.
In the symmetric phase,  one  finds q-balls \cite{coleman}
for some $U(f)$,  a nontopological soliton of net global $U(1)$ charge.
In the broken phase, $\langle T\rangle \neq 0$, the soliton that one
obtains is a global vortex, for which the periodic variable $\theta$ has
a net winding number along some asymptotic circle. Since a continuous
symmetry is broken, a Goldstone boson must be present and is identified
with $\theta$.

When the gauge coupling $e$ is turned on, this global vortex becomes a soliton
of finite energy that also
carries a quantized magnetic flux: The gauge particle becomes massive
by eating up the would-be Goldstone boson $\theta$. Because of this mass
term, the asymptotic behavior of $A_\mu$ is now tied to that of
$\partial_\mu\theta$, leading to a localized magnetic flux whenever
$\theta$ has a winding number. Magnetic flux is quantized as
\begin{equation}
\int_{R^2}  {F} = \oint {A}= \frac{1}{e}
\oint \partial\theta=\frac{ 2\pi n }{e} .
\end{equation}
The asymptotic form of $A$ is clearly a
pure gauge, so the magnetic flux $F$ is localized near $|T|=0$. When we
identify $A$ as the relative $U(1)$ gauge field $A_{12}$, this vortex
soliton in the broken  phase is exactly what Sen identified
as the D0 brane obtained from a D2-anti-D2 pair.

One can ask if there is an analogous effective description when charged string
condenses? First of all, 1-form gauge field $A_\mu$ is promoted to 2-form 
gauge field $B_{MN}$, so naturally the would-be Goldstone boson $\theta$ 
is also promoted to a 1-form, $G_M$. Denote their field strengths by
$H_{MNP}$ and $K_{MN}$, respectively. One can imagine that the relevant
string field is defined over a loop space and that the 1-form $G$ comes from 
a factor of it,
\begin{equation}
e^{i\oint G} .
\end{equation}
The phase factor  $exp(i\oint G)$
must be single-valued in the loop space, and this
quantizes possible net $K$ flux along any closed surface $\Sigma$;
\begin{equation}
\oint_{\Sigma}K={2\pi n} ,
\end{equation}
where $n$ is an integer.
One possible generalization of the effective Lagrangian above is then,
\begin{equation}
{\cal L} = \frac{1}{12}(H_{MNP})^2 +  \frac{1}{2}(\partial_M f)^2
- m(f)^2 (K_{MN} - e B_{MN})^2 - U(f)  ,
\end{equation}
where $f$ is a macroscopic version of some modulus field associated
with the factor $e^{i\oint G}$.
$U$ and $m$ are functions of $f$ that must be determined microscopically.
We wish to argue that this Lagrangian contains many qualitative features
one would expect from string condensation.

Suppose $e=0$ for the moment. The physics of the
symmetric phase is that of a strongly interacting QED with ``gauge field''
$G_M$, coupled to a real scalar $f$. Again, the symmetric phase 
($\langle f\rangle\neq 0$) may contain nontopological soliton which now 
consists of a flux tube of $K_{MN}$, given an appropriate form of $U(f)$.
The existence argument is almost identical to the q-ball
case~\cite{kimyeong}.\footnote{Suppose that the electric field is
spread evenly on transverse volume $V$ and that the scalar field is
constant $f=v$ in this region. The total flux is approximately $\Phi=
f^2 E_{01} V$. The total energy per unit length of the string will be then
\begin{equation}
{\cal E} = \frac{\Phi^2}{2v^2 V} + VU(f) .
\end{equation}
When $V^2= \Phi^2/(2f^2 U(f))$, the minimum of this energy density would be
obtained with value
\begin{equation}
{\cal E} = \Phi \sqrt{\frac{2 U(f) }{f^2}}  .
\end{equation}
Once there is the minimum of $2U(f)/f^2$ away from the symmetric
phase, there exists a stable q-bundle.}  In other words, the theory
contains a solitonic string in the symmetric phase, which couples to
$B$ as a fundamental string would when we turn $e$ back on. This is 
one justification for using the above as a toy model for the string
condensation.

Let us keep $e\neq 0$. When the potential $U$ becomes tachyonic,
the theory settles down in the broken phase, $\langle f\rangle \neq 0$.
Since the would-be Goldstone boson is now a vector, a possible topological
defect is that of a Dirac monopole. The defect imposes a boundary condition at 
infinity of type,
\begin{equation}
\oint_{S^2} K =2\pi n .
\end{equation}
For $n\neq 0$, the 1-form field $G_M$ requires more 
than one coordinate patches, and, if we associate $G$ with a line bundle on
$S^2$, the integer $n$ would be the first Chern class of that line bundle.

In this asymmetric phase, the gauge-invariant mass term for $B$ tells us that
the asymptotic behavior of $B$ is tied to that of $K$.
The flux of $H$ is then quantized as
\begin{equation}
\int_{R^3}  H= \oint_{S^2}B =\frac{1}{e}
\oint_{S^2} K = \frac{2\pi n }{e} .
\end{equation}
The asymptotic form of $B\simeq K$ is pure gauge, since $dK=0$, so the $H=dB$ 
flux is again localized near $f=0$, just as in the vortex soliton, except that
now the resulting solitonic object is of co-dimension three. Given the 
topological coupling (\ref{ch}), this soliton is interpreted as M2 brane.

One may worry whether a finite mass soliton is possible in such
a theory, let alone whether the soliton has the right mass to be a M2 brane.
While one does not know the detailed form of the soliton, it is relatively easy
to see how possible divergences in energy density are avoided. At large
distances, it is the behavior of $H^2$ that could be troublesome. However,
the asymptotics, $B\sim K$, tells us that $H$ decays as $1/r^3$, so the
energy density contribution $H^2$ scales as $1/r^6$. In $(5+1)$ dimensions,
the integrated energy is convergent. Near the soliton cores, where the
match between $B$ and $K$ need not hold any more, it is vanishing $f$ and
thus $m(f)$ that should come to the rescue. By having $m(f)\rightarrow
0$ sufficiently fast, we do find a finite core energy. For instance,
$m(f)\sim f^2$ near $f=0$ will do. (Although, since this effective Lagrangian 
governs long distance dynamics at best, it is not meaningful to ask 
exactly what $m(f)$ is for small $f$. A better prescription might be
to introduce a short distance cut-off.)

\section{Fundamental Strings from D$p$-anti-D$p$; Confinement of $A_S$}

Discussions in previous sections apply to other cases. As we observed
above, NS5 brane of type IIA theory has essentially the same low energy 
behavior  as M5 brane, so the same analysis will show that D2 branes are
produced when an NS5 brane annihilates an anti-NS5 brane. Another way
to see this is to compactify M theory such that neither the M5's nor the 
resulting M2 is wrapped on $S^1$.

One can consider an alternate compactification of M5-anti-M5 to a string 
theory configuration, in such a way that the resulting M2 brane is wrapped on
$S^1$ and becomes a fundamental string.  This M2 would be lying on M5-anti-M5 
world-volume, so the M5's also have to be wrapped on $S^1$ and 
become D4's. It follows that fundamental 
strings are produced when a D4 brane annihilates an anti-D4 brane. One 
naturally wonders how such co-dimension-three objects are possible 
when D4 brane theory is an ordinary gauge theory.

Expanding Born-Infeld action of the D4 brane and the anti-D4 brane, 
one finds terms,
\begin{equation}
\int_{\rm D4} \langle {\cal B}, dA_1\rangle+
\int_{\rm anti-D4}\langle {\cal B}, dA_2\rangle =
\int_{R^{4+1}} \langle {\cal B}, d(A_1+A_2)\rangle , \label{ba}
\end{equation}
where ${\cal B}$ is NS-NS 2-form field pulled back to the common world-volume.
The two gauge fields are summed in this case, because the inner product 
$\langle,\rangle$ is insensitive to the orientation of the world-volume.
Clearly it is the {\it electric} flux of the overall $U(1)$ gauge 
field $A_S\equiv A_1+A_2$ that would generate the fundamental string charge 
of $\cal B$. 

On the other hand, from the above discussions of M5-anti-M5, we know that 
open D2 ending on a D4-anti-D4 pair provides string-like tachyonic objects.
This object is magnetically charged with respect to the world-volume gauge
fields $A_1$ and $A_2$. More specifically, it couples minimally to 
$B'\equiv B_1'-B_2'$ when we dualize the gauge vectors $A_i$'s
to gauge 2-form $B_i'$'s. This actually implies that it
carries a magnetic charge of $A_S$, rather than that of $A_{12}$. 
As we saw in section 3, the Hodge dual operator
on anti-D4 has an extra sign with respect to the one on D4, owing to
the reverse orientation. This introduces a  sign  when we
dualize the gauge field, so we find that
\begin{equation}
d(B_1'-B_2') =*\,d(A_1+A_2) ,
\end{equation}
where the Hodge dual operator $*$ is defined with respect to D4. Thus, 
the tachyonic string is indeed magnetically charged with respect 
to $A_S=A_1+A_2$.

Once this tachyonic string condenses, the $U(1)$ gauge field $A_S$ is
confined by the dual Higgs mechanism. The only remnant of $A_S$ is
the confined electric flux string. The world-volume coupling (\ref{ba})
induces $\cal B$ charge on $A_S$ flux, so the confined flux string
is the fundamental string of type IIA theory.

Let us see how the gauge-invariant mass term for $B'$ arises
from M theory picture. It must be present if dual Higgs mechanism is to
work for the overall $U(1)$. We already noted that, upon 
the compactification on a circle, the M5-anti-M5 world-volume field $B$ 
reduces to a pair of vector fields on D4-anti-D4. Having identified the 
longitudinal part $B_{5\mu}$ as the relative $U(1)$ vector $(A_{12})_\mu$, 
we know that the overall $U(1)$ $A_S$ must be dual to the 
transverse  part  $B_{\mu\nu}$. That is,
$B'$ is nothing but the part of $B$ parallel to the D4-anti-D4 world-volume. 
The would-be Goldstone boson $G_M$ also reduces to a single angle 
$\theta=G_5$ and a vector $a_\mu=G_\mu$. The $B$ mass term (\ref{mass}) 
on M5-anti-M5 is then split into two pieces. The first,
\begin{equation}
((A_{12})_\mu-\partial_\mu\theta)^2 ,
\end{equation}
is the mass term for $(A_{12})$, while the second
\begin{equation}
(B_{\mu\nu}'-\partial_\mu a_\nu+\partial_\nu a_\mu)^2 ,
\end{equation}
is the mass term for the dual photon $B'$. 

This identification also provides an alternate check on the all-important
topological coupling (\ref{ch}) on M5-anti-M5. A dimensional reduction of 
the coupling (\ref{ch}), when 3-form $C_3$ is aligned along $S^1$ and 
becomes NS-NS 2-form $\cal B$, gives us
\begin{equation}
\int_{R^{4+1}} {\cal B}\wedge dB' =
\int_{R^{4+1}} {\cal B}\wedge d(B_1'-B_2')=
\int_{R^{4+1}} \langle {\cal B}, d(A_1+A_2)\rangle   ,\label{bb}
\end{equation}
which is, as we saw in Eq.~(\ref{ba}), part of well-established Born-Infeld 
actions of the D4 brane and of the anti-D4 brane.

Although there is no analogous M theory picture for other D$p$-anti-D$p$
pairs, it is rather clear that a generalization of this mechanism to
arbitrary D$p$-anti-D$p$ must also exist. Open D$(p-2)$ branes ending on the
pair must provide a $(p-2)$-dimensional tachyonic objects, which couples
to the overall $U(1)$ gauge field $A_S$ magnetically. The subsequent
condensation confines the unbroken $U(1)$, and the fundamental string is 
realized as its electric flux string. One example where this is most clearly
seen is the case of D3-anti-D3, where $SL(2,Z)$ U-duality maps fundamental 
strings to D-strings and vice sersa, while keeping D3's invariant.

\section{Summary}

We have observed how the Higgs mechanism of world-volume 2-form field $B$ 
produces co-dimension three solitonic byproducts on M5-anti-M5,
type IIA NS5-anti-NS5, and D4-anti-D4 systems. They are interpreted
as M2 branes, D2 branes, and fundamental strings, respectively.
In particular, the fundamental string that emerges from a D4-anti-D4 pair,
is nothing but the confined electric flux string of the overall $U(1)$
which remains unbroken upon condensation of the open string tachyon.

Generally, there exist a pair of tachyonic objects when a D$p$ brane
is coincident with an anti-D$p$ brane. One is perturbative 
in origin and couples to the relative $U(1)$ electrically, while the other is 
nonperturbative and couples to the overall $U(1)$ magnetically. The 
combined electric and magnetic Higgs phenomena introduce mass gaps to 
both $U(1)$ gauge sectors. The only remnants are D$(p-2)$ branes 
and fundamental strings, which are realized as ``solitons'' on the common 
$(p+1)$-dimensional world-volume. In particular, this neatly addresses
the question raised by Witten \cite{K} of how the unbroken $U(1)$ becomes
invisible on the annihilated D$p$-anti-D$p$ world-volume.

\vskip 1cm
\centerline{\large\bf Acknowledgment}
\vskip 5mm
The author benefited from conversations with several visitors 
to Korea Institute for Advanced Study (KIAS). Kentaro Hori's probing 
questions lead to many of the insights of this paper, while Mans Henningson's 
expertise on M5 brane was very helpful. Most of all, the author is grateful to
Kimyeong Lee who explained how a Higgs mechanism of 2-form tensor should be
treated, and who is largely responsible for content of section 3.
The author also thanks Miao Li for critical comments. This work is
supported in part by National Science Foundation of the United States.

\end{document}